\documentstyle[12pt,a41,epsfig]{article}

\begin{document}

\begin{center}
{\LARGE\bf Semi--inclusive asymmetries with polarized proton beams at HERA}

\vspace{1cm}
{\underline{M.~Maul}$^a$, J.~G.~Contreras$^b$, H.~Ihssen$^{c}$, and
A.~Sch\"afer $^a$}

\vspace*{1cm}
{\it $^a$ Naturwissenschaftliche Falult\"at II, Physik. Universit\"at
Regensburg, Universit\"atsstr.~31, D--93053 Regensburg, Germany.}\\

\vspace*{3mm}
{\it $^b$Universit\"at Dortmund, Institut f\"ur Physik, D--44221
    Dortmund, Germany.}

\vspace*{3mm}
{\it $^c$ Nationaal Instituut voor Kernfysica en Hoge-Energiefysica (NIKHEF),
P.~O.~Box 41882, 1009 DB Amsterdam, The Netherlands.}\\

\vspace*{2cm}

\end{center}
\begin{abstract}
The prospects of semi-inclusive measurements with polarized
proton beams at HERA are discussed. Detailed simulations show
that one can disentangle the valence-quark and sea-quark
contribution to the polarized structure function $g_1(x)$ in
the small $x$-domain, if the equivalent of 1000 pb$^{-1}$ of data
are collected. It is also shown how semi-inclusive charged-
current events can provide information on the relative
importance of the spin contribution of the anti-s and 
anti-d sea quarks. Moreover, various methods to determine
the fragmentation functions in this kinematical domain are
presented.
\end{abstract}

\section{Introduction}
While inclusive measurements are only sensitive to the sum of all
quark flavors weighted by the square of their charge, 
semi-inclusive measurements allow to disentangle the valence-quark and
sea-quark contributions to structure functions \cite{semii}. 
Moreover they allow for
the measurement of fragmentation functions, which are sensitive to the 
nature of final state interactions.
\newline 
\newline
Fragmentation functions and parton distributions cannot be determined
at the same time, as there are too many unknowns in the cross section.
But one can make important cross checks by fixing the parton distributions
by common parameterizations and determining subsequently the fragmentation
functions.
\newline
We compare three different methods of extracting fragmentation functions
in section \ref{sec2} and
show to which extend these methods allow to disentangle the flavor 
structure of the nucleon. In section  \ref{sec3}  semi-inclusive
measurements of pure $\gamma$-exchange events at small $x$ and $Q^2$ are
discussed. We
show that with polarized proton beams at HERA 
interesting information on the ratio of
valence- and sea-quark distributions can be obtained. In the last section
charged current (CC) events at high $Q^2$ are discussed, where semi-inclusive measurements
 might give some very useful signature in case the strange- or 
the down-sea should be unexpectedly large.
\newline
\newline
\vspace{1mm}
\noindent
\section{Fragmentation functions}
\label{sec2}
\begin{figure}
\centerline{\psfig{figure=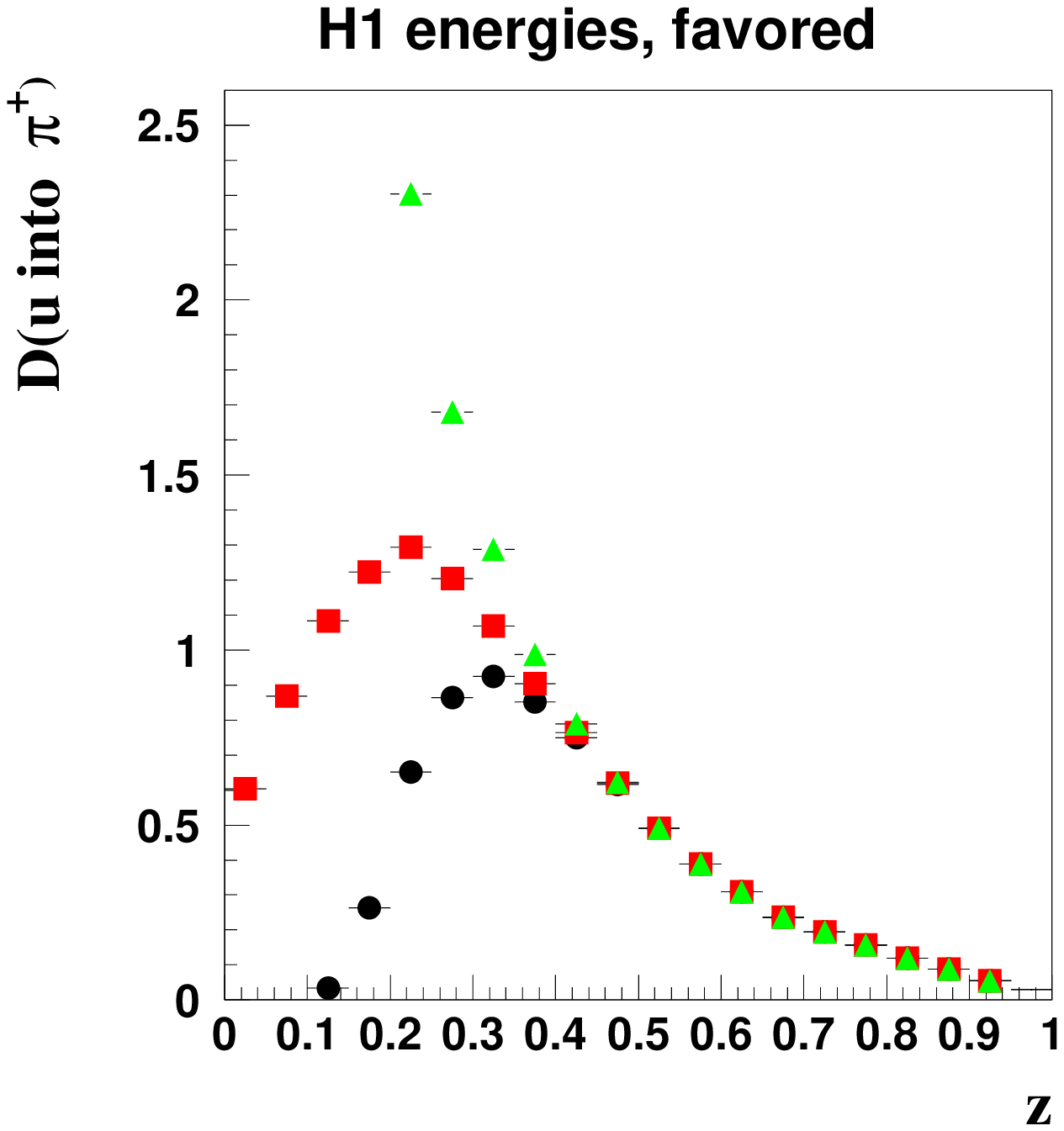,width=5cm}
\psfig{figure=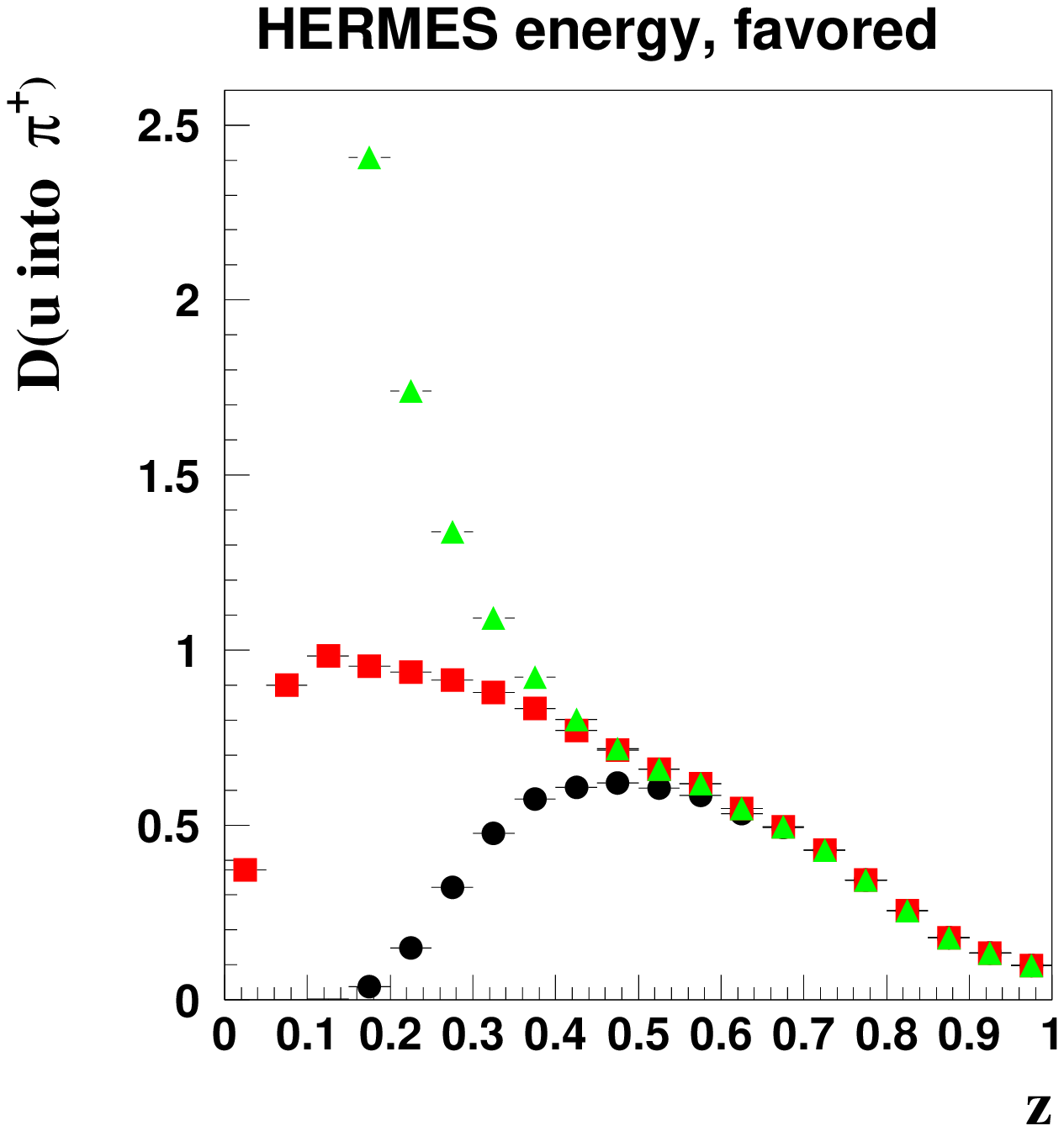,width=5cm}}
\centerline{\psfig{figure=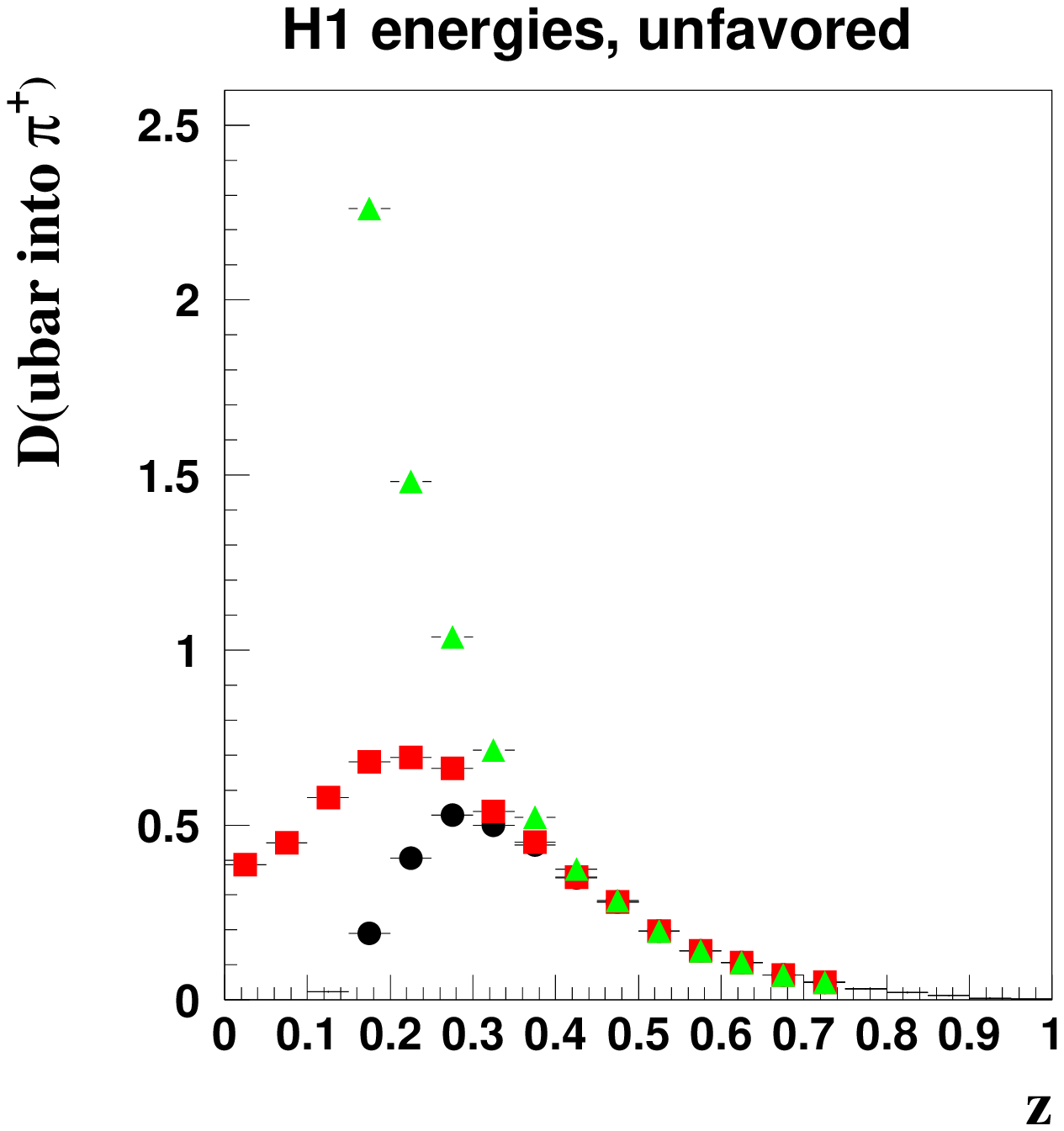,width=5cm}
\psfig{figure=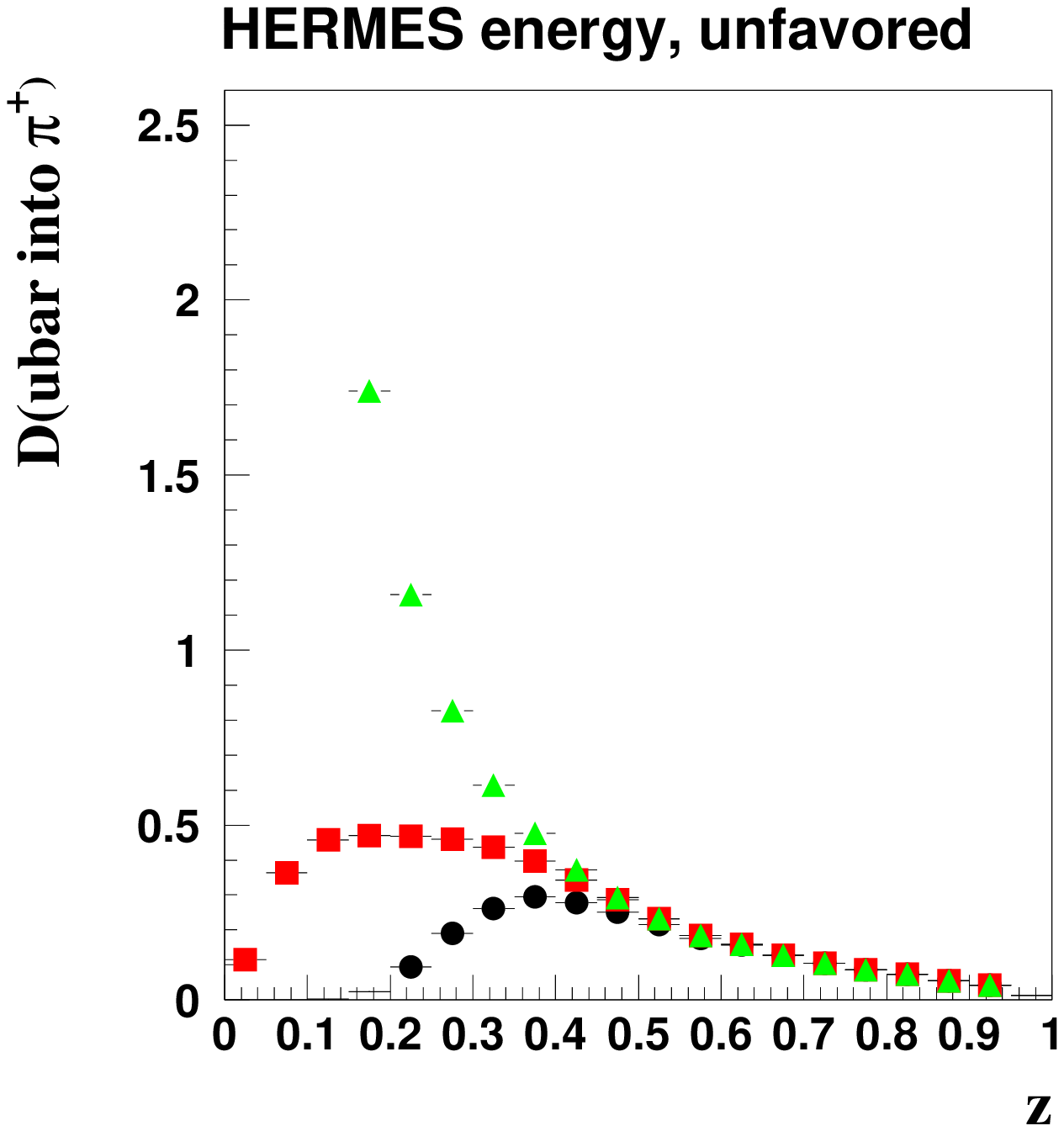,width=5cm}}
\caption{Comparison of the three different extraction methods: triangles -
inclusive method, squares -  pion/kaon method, circles - maximum method.
Ideal $4\pi$ simulation.}
\label{comp}
\end{figure}

\begin{figure}
\centerline{\psfig{figure=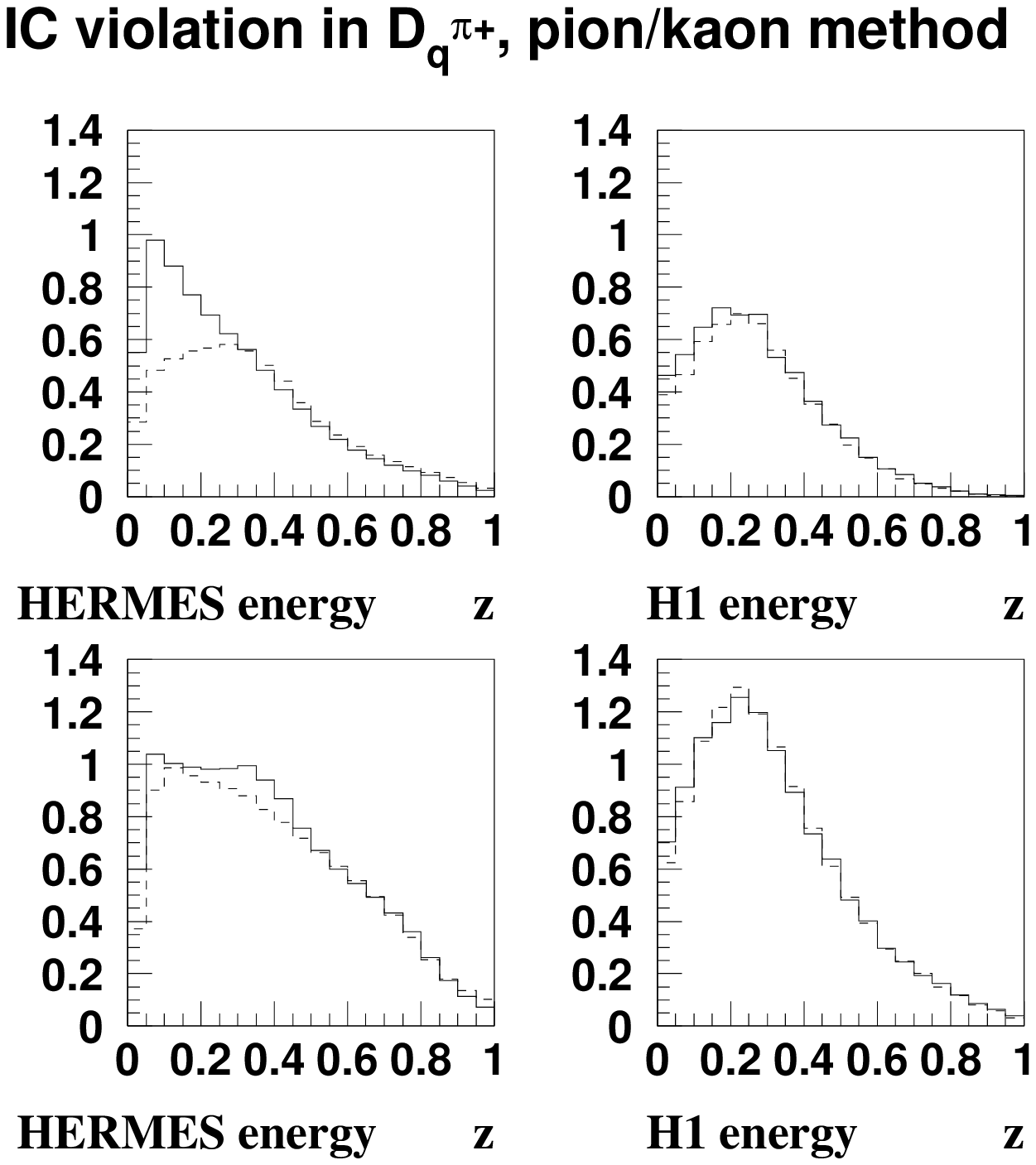,width=10cm}}
\caption{Energy dependence of the IC-violation with respect to
HERMES energies and H1 energies. Ideal $4\pi$ simulation. 
Above: solid line 
$D_{d}^{\pi+}$, dashed line $D_{\bar u}^{\pi+}$; below: solid
line $D_{\bar d}^{\pi+}$ , dashed line $D_{u}^{\pi+}$. }
\label{iso}
\end{figure}

\begin{figure}
\centerline{\psfig{figure=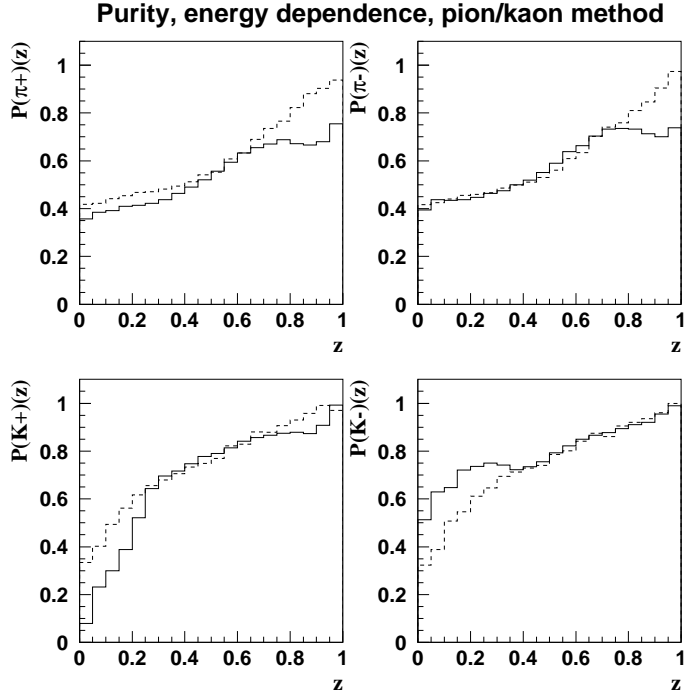,width=10cm}}
\caption{Energy dependence of the purity for
HERMES energies (solid line) and H1 energies (dashed line).
Ideal $4\pi$ simulation.}
\label{pur}
\end{figure}

The primary goal of semi-inclusive measurements is to identify those
particles (leading particles) which contain a quark which has been
struck by the incoming photon or weak gauge boson. 
Generally, for the semi-inclusive cross section the following
decomposition is chosen which defines the fragmentation functions $D_q^h$:
\begin{eqnarray}
\frac{d \sigma_{\rm unpol.}}{dx d{Q^2} dz}
&=& \frac{2\pi \alpha^2_{\rm em} }{Q^4} \sum_{f} e_f^2 ( 1+(1-y)^2) 
q_f(x) D_q^h(z)
\nonumber \\
\nonumber \\
\frac{d \sigma_{\rm pol.}}{dx d{Q^2} dz}
&=& \frac{2\pi \alpha^2_{\rm em} }{Q^4} \sum_{f} e_f^2 ( 1-(1-y)^2) 
 \Delta q_f(x) D_q^h(z)\quad,
\end{eqnarray}
where the unpolarized and polarized fragmentation functions are supposed to
be equal. 
(Note that this assumption is certainly violated at some level. The size of 
such violations can be determined experimentally by e.g.~analyzing 
the $x$-dependences for fixed target experiments.) 
$x$ is the usual Bjorken-x, $y = P\cdot q/(P\cdot k)$,
$ z = P\cdot P_h/(P\cdot q)$
with $P$ being the momentum of the incoming nucleon, 
$k$ of the incoming lepton,
$q$ of the exchanged photon, and $P_h$ of the measured hadron $h$. $q_f, \Delta
q_f$ denote the unpolarized and polarized parton densities with flavor $f$.
$Q^2=-q^2$.
\newline 
\newline
In general for extracting parton distributions from semi-inclusive
measurements one normally identifies the favored and unfavored
fragmentation functions:
\begin{eqnarray}
\label{fragm}
D_{     u}^{\pi+}(z) =  D_{\bar d}^{\pi+}(z) &=& 
D_{\bar u}^{\pi-}(z )=  D_{     d}^{\pi-}(z)  \quad {\rm (favored)} 
\nonumber \\
D_{     u}^{\pi-}(z) =  D_{\bar d}^{\pi-}(z) &=& 
D_{\bar u}^{\pi+}(z )=  D_{     d}^{\pi+}(z) 
\nonumber \\
= D_{     s}^{\pi+}(z) =  D_{\bar s}^{\pi+}(z) &=& 
D_{\bar s}^{\pi-}(z )=  D_{     s}^{\pi-}(z) \quad {\rm (unfavored)} \quad. 
\end{eqnarray}
The form of the fragmentation function depends on the way the hadron $h$ 
is measured. Therefore in principle no normalization independent of the
extraction procedure is possible. Usually, the normalization is fixed 
by the second moment $\int_0^1 z D(z)dz$. As we will
discuss three completely different ways of isolating leading particles, we
base our normalization simply on the cross section formula, this is
reasonable because one of the discussed extraction procedures allows
for more than one 'leading particle' per event, while the two others do not.
The following three extraction methods are regarded:
\begin{itemize}
\item {\bf \underline{Inclusive method:}} 
All final state particles are taken into account.
\item{\bf \underline{Pion}/\underline{kaon method:}} 
The leading particle is the charged pion or kaon with the maximum $z$ of
all charged pions or kaons.
\item{\bf \underline{Maximum method:}} 
The leading particle is the final state particle
which carries the maximum $z$ of all final state particles (including 
photons and baryons).
\end{itemize}
The pion/kaon method is inspired by the fact that leading
particles are mostly mesons. For our Monte-Carlo simulations we use
the program {\sc Lepto-6.5} \cite{IER97} for event generation 
together with {\sc Jetset-7.4}
\cite{S94} for hadronization. \newline
\newline
In Fig.~\ref{comp}  the three methods of extraction are compared with 
each other in a $4\pi$ simulation without detector acceptances in order
to concentrate on the energy dependence. 
For $z>0.5$ all methods result in the same values
as to be expected by energy conservation,
but the inclusive method has by far the largest multiplicity, while
the maximum method gives no particles for $z<0.2$ for HERMES and $z<0.1$ for
H1 energies. 
\newline
Fig.~\ref{iso} shows in an ideal $4 \pi$ simulation 
how well the assumed combined 
isospin and charge conjugation (IC) symmetry, 
i.e $D_u^{\pi+} = D_{\bar d}^{\pi+}$ in the favored case and 
$D_{\bar u}^{\pi+} = D_{\bar d}^{\pi+}$ in the unfavored case, is
fulfilled. Within
the Lund string fragmentation model with its standard settings we
observe deviations from this IC-symmetry assumption only
in the small $z$ region for HERMES energies which might be a hint
that the identification of favored and unfavored fragmentation
functions is for HERA-energies even better justified than for
experiments with lower energies.
\newline
\newline
Another very essential question is to ask whether an 
extracted particle is really a leading particle, i.e
that it contains a struck quark or at least a quark which has the
same flavor as the quark hit by the incoming photon. For this
purpose we regard the ratio of the favored fragmentation functions
of a hadron $h$ to all fragmentation functions of the same hadron 
and define this quantity as purity $P_h$
\begin{equation}
P_h = \frac{\sum_{q \in {\rm favored}}D_q^h}
           {\sum_q D_q^h} \quad.
\end{equation}
There exist deviating definitions of purity \cite{Hughes}, but
all agree in so far as for purity of 1 all
fragmentation processes are favored ones. The smaller the purity, 
the less reliable is the identification of leading particles.
In Fig.~\ref{pur} we plot the purity for HERMES and H1 energies 
again in an ideal $4\pi$ simulation
for charged pions and kaons. For pions the purity for small $z$ is
nearly independent of the energy. For large $z$ it reaches up to 90\%
for H1 energies, but only up to 70\% for lower energies. 
For the kaonic purities for the HERMES energies a charge conjugated
anomaly is visible which changes its sign from $K^+ \longrightarrow K^-$.
\newline
\newline
In summary we find that with the H1 energies  deviations from IC symmetry
are negligible and that the identification in favored and unfavored 
fragmentation functions is better fulfilled than for lower energies.
\section{Semi-inclusive asymmetries for $\gamma$-exchange at
small $x$}
\label{sec3}
\begin{figure}
\centerline{\psfig{figure=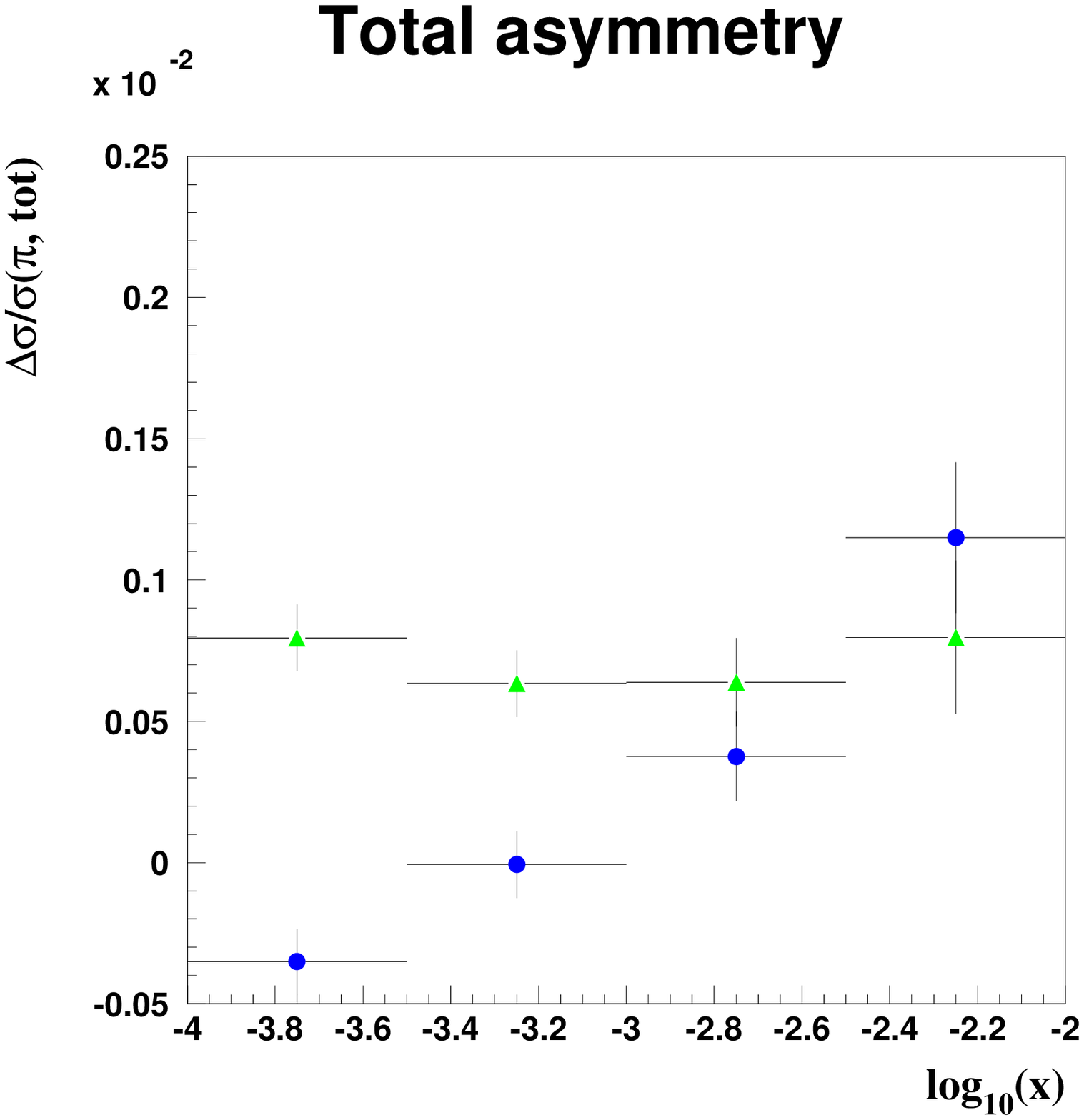,width=9cm}}
\caption{Total pionic asymmetry 
$\frac{1}{P_e P_p}\frac{\Delta \sigma^{\pi+}+ \Delta \sigma^{\pi-}}  
{\sigma^{\pi+}+ \sigma^{\pi-}}$ for 500 pb${^{-1}}$ per
relative polarization and polarization degree $P_e = P_p = 0.7$.
Triangles GSA, circles GRSV LO STD.
Simulation done with {\sc Pepsi}.}
\label{tot}
\end{figure}

\begin{figure}
\centerline{\psfig{figure=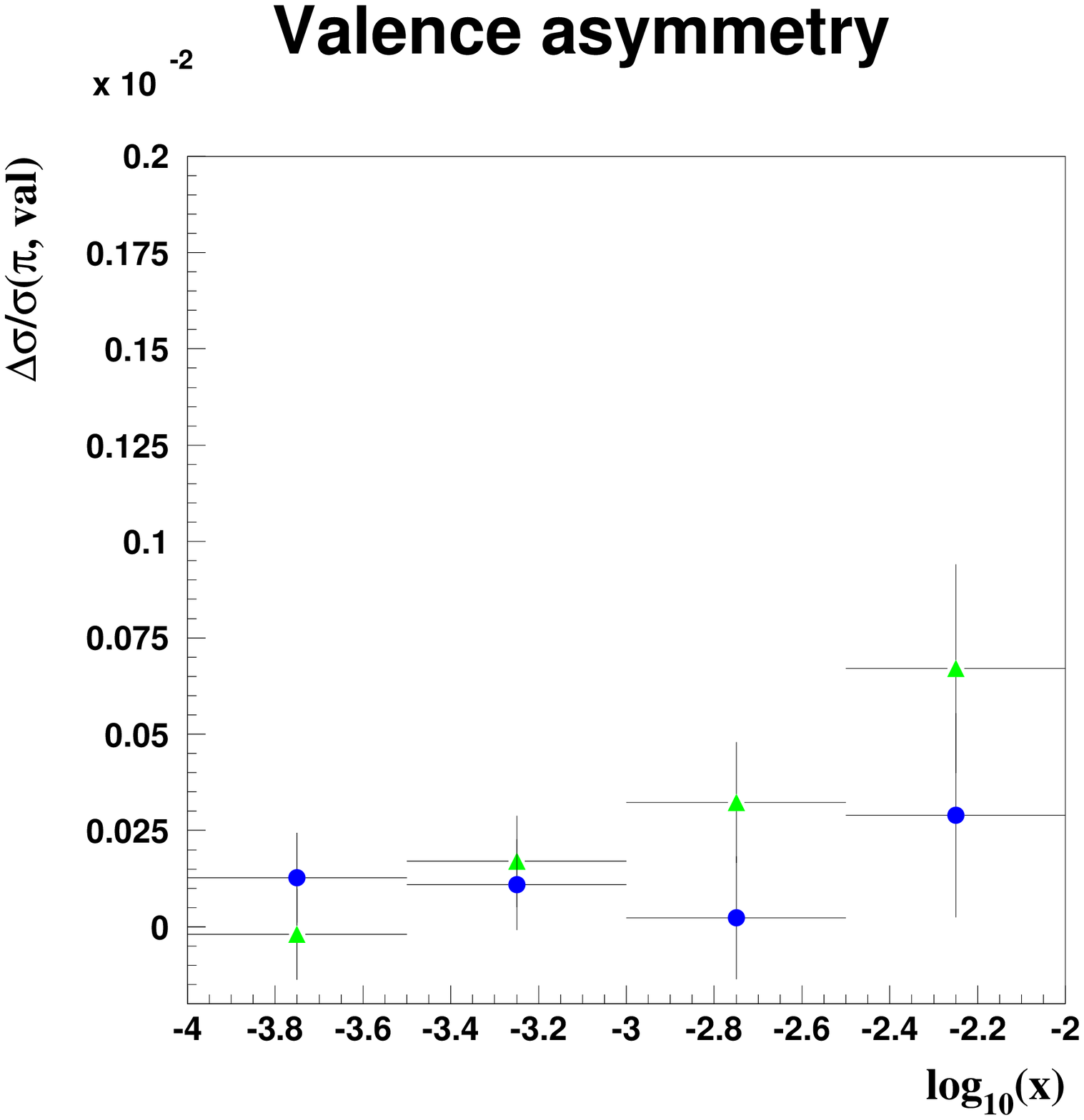,width=9cm}}
\caption{Valence sensitive pionic asymmetry 
$\frac{1}{P_e P_p}\frac{\Delta \sigma^{\pi+}- \Delta \sigma^{\pi-}}
{\sigma^{\pi+}+ \sigma^{\pi-}}$ for 500 pb${^{-1}}$ per
relative polarization and  polarization degree $P_e = P_p = 0.7$.
Triangles GSA, circles GRSV LO STD.
Simulation done with {\sc Pepsi}.}  
\label{val}
\end{figure}

At small $x$, in the unpolarized case, the sea is dominating over the
valence quarks. Polarized proton beams at HERA  
would allow to determine additional combinations of valence and  sea 
quark distributions. The following two asymmetry combinations
are of interest:
\begin{eqnarray}
\label{asym}
\Delta \sigma/\sigma(\pi,val) &:=& \frac{\Delta N(\pi^+) - \Delta N(\pi^-)}
                                {\Sigma N(\pi^+) + \Sigma N(\pi^-)}
= {\cal D} \frac{ (4 \Delta u_v - \Delta d_v)(D-\overline D)}
       { (4 u_{\rm tot} + d_{\rm tot})(D+\overline D) + 2s\overline D}
\;,
\nonumber \\
\nonumber \\
\Delta \sigma/\sigma(\pi,tot) &:=& \frac{\Delta N(\pi^+) + \Delta N(\pi^-)}
                                {\Sigma N(\pi^+) + \Sigma N(\pi^-)}
=  {\cal D} 
\frac{ (4 \Delta u_{\rm tot} + \Delta d_{\rm tot})(D+\overline D)
        +2 \Delta s\overline D}
       { (4 u_{\rm tot} + d_{\rm tot})(D+\overline D) + 2s\overline D}
\;.
\nonumber 
\end{eqnarray}
$D,\overline D$ are the favored and unfavored fragmentation functions
as defined in Eq.~\ref{fragm}.
${\cal D}$ is the depolarization factor, $\Delta N(\pi^+), \Sigma N(\pi^+)$
the difference and the sum, respectively, of the $\pi^+$ multiplicities for
the two different spin configurations. In the numerator we consider the
sum and the difference of the polarized multiplicities $\Delta N (\pi^+)$
and $\Delta N (\pi^-)$. The sum of both contains
contributions from the valence and the sea quarks,
while the difference is only sensitive to the valence-quark distributions.
For the unpolarized parton distributions at small $x$, the
valence contribution is small, namely on the percent level. Consequently,
in the denominators of Eq.~\ref{asym} only sea-quark 
parton distributions contribute effectively. 
\newline
\newline
For pions  we choose in order to maximize
the luminosity the inclusive extraction method, but require $z>0.2$. We make
no further restrictions than those which are given by the
limitations of the H1 detector and the HERA itself 
($E_p$ = 820 GeV, $E_e$ = 27.5 GeV) :
\begin{eqnarray}
\label{param}
0.01 < &y& < 0.9 \nonumber \\
0.0001 < &x& < 0.01 \nonumber \\
177 \;{\rm deg}\; < &\Theta_e^{\rm  (scat)} \nonumber \\
5\; {\rm GeV} < & E_e ^{\rm (scat)} \nonumber \\
 0.15 \;{\rm GeV/c} < &p_{\perp \pi,K}^{\rm track,lab}&   \nonumber \\
20 \;{\rm deg.}\;  < &\Theta_{\pi,K}^{\rm track,lab}& < 150 \;{\rm deg.}\; 
\qquad. 
\end{eqnarray}
For our polarized MC simulations we use a new version of
{\sc Pepsi} \cite{pepsi1}.
Presently one can only speculate about the behavior of $g_1$
at small $x$. Gehrmann/Stirling LO Gluon A (GSA) \cite{GS95} for example,
assumes $g_1$ for H1 energies to be positive, i.e a positive
sea-quark polarization at small $x$, while the opposite assumption
(negative $g_1$ at small $x$) is made by GSRV LO STD \cite{GRSV95}. 
Both parameterizations use the LO set \cite{GRV94} for 
the unpolarized parton distributions.
This behavior is well reflected in the semi-inclusive asymmetry
$\Delta \sigma/\sigma(\pi,tot)$  
in Fig.~\ref{tot}, where at $x < 10^{-3.5}$ the simulation
for the GRSV LO STD shows a negative value while the corresponding
MC-values for the Gehrmann/Stirling set still remain positive. 
It would be interesting to check experimentally how  the
sea and valence distributions behave separately. This can 
be determined experimentally by measuring the valence asymmetry 
$\Delta \sigma/\sigma(\pi,val)$ (see
Fig.~\ref{val}). For both sets the valence contribution is of the
same size within the error bars. It becomes small for decreasing
$x$ and points to a dominant sea at small $x$. 
If, e.g., the valence asymmetries would show nearly the same behavior
as the total asymmetries $\Delta \sigma/\sigma(\pi,tot)$  
then the polarized valence contribution
would be dominant.
\newline
\newline
In this way semi-inclusive measurements provide an important
tool to disentangle at small $x$ the valence and the
sea contributions. These measurements can only be done in the foreseeable
future with polarized proton beams at HERA. 
\vspace{1mm}
\noindent
\section{Semi-inclusive measurements for $W^-$ exchange}
\label{sec4}
\begin{figure}
\centerline{\psfig{figure=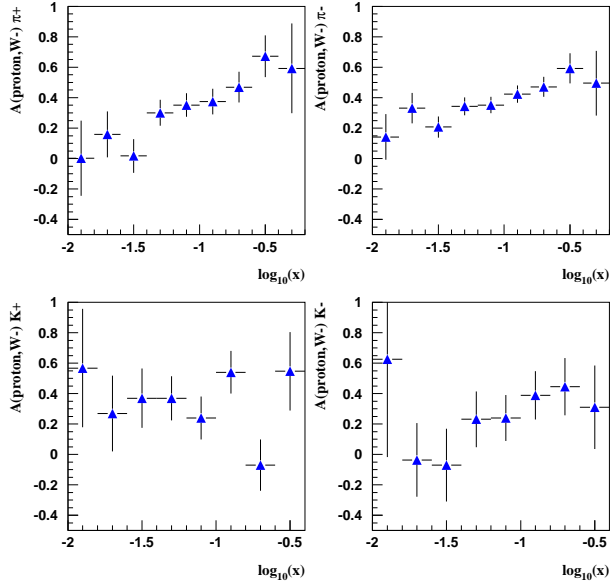,width=9cm}}
\caption{Semi-inclusive asymmetries 
$\frac{1}{P_e P_p}\frac{\Delta \sigma^{H}}
{\sigma^{H}}, H= \pi^+,\pi^-,K^+,K^-$ for $W^-$ exchange, for 200 pb${^{-1}}$
per relative polarization and  polarization degree $P_e = P_p = 0.7$.
Parton distribution set GRSV LO STD.
Simulation done with {\sc Pepsi}.}  
\label{ewm}
\end{figure}

While semi-inclusive pionic asymmetries allow to separate
the valence and the sea contributions to structure functions,
one can distinguish positively from negatively charged flavors
via $W^\pm$ exchange. At $Q^2 > 600$ GeV${^2}$ those
events are a frequent subprocess in DIS. They can
be identified by the missing momentum.
In the semi-inclusive case 
we deal with asymmetries of the form 
\begin{eqnarray}
{\cal A}(W^-,h) &=& 
\frac{1}{P_e P_p} \frac{\Delta \sigma^h}{\sigma^h}\Bigg|_{W^-}
= \frac{\Delta u D_u^{h,W} - (y-1)^2(\Delta \bar s D_{\bar s}^{h,W}  
    + \Delta \bar d D_{\bar d}^{h,W})}
{ u D_u^{h,W} + (y-1)^2( \bar s D_{\bar s}^{h,W}
    +  \bar d D_{\bar d}^{h,W})}\quad,
\end{eqnarray}
where $h$ is an outgoing hadron. $D_q^{h,W}$ describes the
charged current fragmentation function. As opposed to the 
neutral current exchange, the struck quark changes its flavor to
a flavor with opposite isospin within the same generation, 
except for flavor mixing.
The identification $D_{\bar s}^{h,W} \equiv D_{\bar c}^h$ is plausible, but
may be wrong because of kinematic differences.
\newline
\newline
For given parton distributions, 
one can analyze these asymmetries with regard to
quark flavor:
\begin{itemize}                     
\item
$W^-$ triggers only on positively charged flavors.
\item
${\cal A}(W^-,\pi^+)$, ${\cal A}(W^-,\pi^-)$ differ only with respect to 
the anti-strange quark contribution.
\item
${\cal A}(W^-,K^+)$, ${\cal A}(W^-,K^-)$ differ only with respect to
the anti-down quark contribution.
\end{itemize}
The pionic and kaonic asymmetries are given in Fig.~\ref{ewm}. 
The same machine parameters are used as in the previous section (see
Eq. \ref{param}).
The corresponding asymmetries for 
the $W^+$ case are unfortunately much smaller.
The error bars are given for a luminosity 
of 200 pb${^{-1}}$ per relative polarization 
and polarization degree of 70\% for the electron
and the proton beam each. As the $\bar s$ and $\bar d$ contributions
are small compared to the u-quark distribution, in our simulation
the pionic and kaonic asymmetries are nearly equal.
But if  $\bar s$ and $\bar d$ would be much larger than assumed in 
the parameterization of GRSV 96 STD LO, one would recognize significant
deviations.

\section{Conclusions}
Semi-inclusive measurements at an upgraded 
HERA with polarized proton beams and a luminosity of 1000 pb${^{-1}}$ 
can provide a key for understanding the valence and sea behavior at small $x$.
Important cross checks for the u-quark dominance over
$\bar s$ and $\bar d$ distributions can be done with polarized CC
experiments. Moreover, polarized proton beams at HERA  will offer
the possibility to test several assumptions and models of fragmentation 
functions.


\end{document}